\documentclass[epj]{svjour}
\usepackage{times}  
\usepackage{epsfig}
\usepackage{fancybox}
\usepackage{amsmath}
\allowdisplaybreaks[2]
\newcommand{\be}{\begin{equation}}
\newcommand{\en}{\end{equation}}
\newcommand{\bea}{\begin{eqnarray}}
\newcommand{\ena}{\end{eqnarray}}
\newcommand{\lbl}[1]{\label{eq:#1}}
\newcommand{\lbltab}[1]{\label{tab:#1}}
\newcommand{\lblfig}[1]{\label{fig:#1}}
\newcommand{\lblsec}[1]{\label{sec:#1}}
\newcommand{\rf}[1]{(\ref{eq:#1})}
\newcommand{\Table}[1]{\ref{tab:#1}}
\newcommand{\fig}[1]{\ref{fig:#1}}
\newcommand{\sect}[1]{\ref{sec:#1}}
\newcommand{\dslash}{%
\mathrel{\setbox0=\hbox{$\partial$}\copy0\kern-0.8\wd0\hbox{\slash}}}
\newcommand{\Aslash}{%
\mathrel{\setbox0=\hbox{$F$}\copy0\kern-0.8\wd0\hbox{\slash}}}
\newcommand{\pslash}{%
\mathrel{\setbox0=\hbox{$p$}\copy0\kern-0.8\wd0\hbox{\slash}}}
\newcommand{\kslash}{%
\mathrel{\setbox0=\hbox{$k$}\copy0\kern-0.8\wd0\hbox{\slash}}}
\newcommand{\qslash}{%
\mathrel{\setbox0=\hbox{$q$}\copy0\kern-0.8\wd0\hbox{\slash}}}
\newcommand{\qqslash}{%
\mathrel{\setbox0=\hbox{$q'$}\copy0\kern-1.0\wd0\hbox{\slash}\kern0.5\wd0}}
\newcommand{\gapprox}{%
\mathrel{%
\setbox0=\hbox{$>$}\raise0.6ex\copy0\kern-\wd0\lower0.65ex\hbox{$\sim$}}}
\newcommand{\lapprox}{%
\mathrel{%
\setbox0=\hbox{$<$}\raise0.6ex\copy0\kern-\wd0\lower0.65ex\hbox{$\sim$}}}
\newcommand{\MeV}{{\rm MeV}}

\newcommand{\mpi}{m_\pi}
\newcommand{\mk} {m_K}
\newcommand{\mpid}{{m_\pi^2}}
\newcommand{\mpit}{{m_\pi^3}}
\newcommand{\mpiq}{{m_\pi^4}}
\newcommand{\fpid}{{F_\pi^2}}
\newcommand{\mpd}{{m_+^2}}
\newcommand{\mmd}{{m_-^2}}
\newcommand{\mkd}{{m_K^2}}
\newcommand{\mud}{{\mu^2}}
\newcommand{\metad}{m_\eta^2}
\newcommand{\mtaud} {{M_\tau^2}}
\newcommand{\Kbar} {\overline{K}}
\newcommand{\trace}[1]{\langle #1 \rangle}
\newcommand{\Chip}{\chi_+}
\newcommand{\Chim}{\chi_-}
\def\lag{ {\cal L} }

\def\mv {{M_V}}
\def\mvd{{M_V^2}}
\def\mvq{{M_V^4}}
\def\gvd{{G_V^2}}

\def\gvun{   {\, g^m_{V1}} }
\def\gvdeux{ {\, g^m_{V2}} }
\def\ghvun{   {\,\hat g^m_{V1}} }
\def\ghvdeux{ {\,\hat g^m_{V2}} }
\def\fvun{   {\, f^m_{V1}} }
\def\fvdeux{ {\, f^m_{V2}} }
\def\evm{ {e_V^m}}
\def\fchi{ {f_\chi}}
\def\mhat{\hat{m}}
\def\ms{ {m_s}}
\def\mdiff{{(m_u-m_d)}}

\def\Kbar{ {\overline{K}} }

\begin{document}

\title{ Tests of the naturalness of the coupling constants \\
in ChPT at order $p^6$}

\author{
K. Kampf\inst{1}
\and
B. Moussallam\inst{2}
}

\institute{Institute of Particle and Nuclear Physics, Charles University
V Hole\v{s}ovi\v{c}k\'ach 2, \\
CZ-180 00 Prague 8, Czech Republic
\and
Institut de Physique Nucl\'eaire,
Universit\'e Paris-Sud 11, F-91406 Orsay, France}

\date{}

\abstract{
We derive constraints on combinations of $O(p^6)$ chiral coupling
constants by matching a recent two-loop calculation of the $\pi K$ scattering
amplitude with a set of sum rules.
We examine the validity of the natural expectation that the values of the
chiral couplings can be associated with physics properties of the light
resonance sector. We focus, in particular, on flavour symmetry breaking
of vector resonances.
A resonance chiral Lagrangian is constructed which incorporates
flavour symmetry breaking more completely than was done before. We
use $\pi K$ unsubtracted sum rules as tests of the modelling  of
the resonance contributions to the chiral couplings.
In some cases the $O(p^6)$ couplings are found not to be dominated by
the resonance contributions.
} 


\maketitle

\section{Introduction}

An important progress in the description of QCD via effective theories was
achieved by the extension of the chiral expansion
formalism\cite{weinberg79,gl84,gl85}
to the order $p^6$\cite{fearing,bce99div,bce99class}.
This raises the hope of attaining high precisions in the description of
low energy physics using the chiral expansion, even in the case of the
three flavour expansion which is expected to converge more slowly than the
two flavour one.
A large number of quantities have already been computed at chiral order
six starting from the work of ref.\cite{bellucci}.
Some representative examples concerning the two-flavour case are
in refs.\cite{bcegs,su2p6} and in the three-flavour case
in refs.\cite{golokamb,ABT2pt,Kl3,ABTKl4a,bijdscalff,bijdontpik}.

In practice, including the $O(p^6)$ corrections was shown
to clearly bring significant improvement for the two-flavour
expansion\cite{bellucci,bcegs}. In this case, the corrections are dominated
by the chiral logarithms, the coefficients of which are known in terms of
the $O(p^2)$ and $O(p^4)$ coupling constants\cite{weinberg79}, while
the corrections proportional to the $O(p^6)$ couplings are comparatively
smaller. The situation for the three-flavour expansion is different,
in that the role of the $O(p^6)$ couplings is much more important. As an
example, in order to determine the CKM matrix element
$V_{us}$ at the one percent level
based on experimental data on $K\to\pi l\nu$ decays it is
necessary to know the values of the two LEC's $C_{12}^r$ and
$C_{34}^r$ (see e.g.\cite{cirigkl3}).

As far as only the order of magnitude  of the chiral LEC's
is concerned, it is possible to make
very simple and general statements\cite{georgiman,georgi}.
The order of magnitude can be argued to depend only on $F_\pi$ and on
the chiral scale $\Lambda_\chi\simeq M_\rho$ such that the typical size of the
$O(p^4)$ LEC's should be $L_i^r\sim F_\pi^2/M_\rho^2$ and that of the
$O(p^6)$ LEC's should be $C_i^r\sim F_\pi^2/M_\rho^4$. The natural question
which arises, then, is whether it is possible to make more
quantitative estimates
relating the values of the LEC's to known properties of the light resonances
in the QCD spectrum. A detailed study along this line was performed
in ref. \cite{egpr} in which it was observed that it is indeed possible to
reproduce the values of the $O(p^4)$ LEC's $L_i^r(\mu)$ with $\mu=M_\rho$,
which had previously been determined in a model independent way\cite{gl85},
in terms of observables from the light resonance sector.

A justification for such a relationship is provided by the chiral sum rules
(see e.g.\cite{gl84}  for a list). A typical example, which was analyzed
in refs.\cite{donogol93,daviergir} is the LEC $L_{10}^r$ which can be
expressed as a convergent integral in terms of spectral functions which
can be determined experimentally from $\tau$ decays. To a good approximation,
the integral is found to be saturated by the contributions from the
$\rho(770)$ and $a_1(1230)$ mesons.
In more complicated situations, for which the integrands cannot easily be
measured, one can appeal to the large $N_c$ expansion.
Indeed, at leading order in $1/N_c$ QCD can
be re-expressed in terms of a Lagrangian involving
an infinite set of
weakly interacting mesons\cite{thooft}. The precise form of this
Lagrangian is not yet
known from first principles, but the weak coupling property
allows one to relate the coupling constants  to observables
using tree level calculations, and then deduce the values of these
observables from experiment.

How well does resonance saturation perform in determining the size
of the $O(p^6)$ LEC's $C_i^r$ is not known at present. The main reason
is that very few
of these LEC's have been determined so far. The purpose of this paper is
to derive some constraints on the  LEC's $C_i^r$ obtained by equating
the $\pi K$ scattering amplitude in the subthreshold region, as constructed
from experimental data in ref.\cite{roypik},
with the chiral expansion calculation up to order $p^6$
which was performed in ref.\cite{bijdontpik} (previous work
comparing dispersive representations with the chiral expansion
up to order $p^4$\cite{bkm} was performed in refs.\cite{anantb,anantbm}).
Some of the $\pi K$
subthreshold expansion parameters can be expressed as unsubtracted
sum rules. Such expressions allow one to identify resonance contributions
from experiment. We will use such results to compare with the same resonance
contributions as computed starting from a large $N_c$ type
resonance chiral Lagrangian.
We will concentrate on a set of contributions  arising from vector meson
resonances and which can be related to flavour symmetry breaking  in the
meson multiplet. It is known that the $\pi\pi$ or $\pi K$
scattering amplitudes receive comparable contributions from vector mesons and
from scalar mesons. Describing scalar mesons starting from a resonance
chiral Lagrangian presents several difficulties, notably in identifying
the properties of the nonet in the chiral limit and in the treatment of the
wide resonances.  For this reason, we will concentrate here on the vector
resonances.

The plan of the paper is as follows. We start by recalling some notation
concerning the $\pi K$ scattering amplitude and some aspects of the
correspondence between the expansion parameters around the subthreshold
point $t=0,\ s-u=0$ and the $O(p^6)$ LEC's. Results concerning the LEC's
$C^r_1$ to $C^r_4$ (which are associated with six derivatives chiral
operators) are then presented. We next consider $\pi K$ subthreshold parameters
associated with chiral operators involving four derivatives plus one
quark mass matrix.
In this sector, serious discrepancies are observed
between the chiral predictions and the sum rule results.
We point out some deficiencies of the resonance model employed
in ref.\cite{bijdontpik} for the relevant $O(p^6)$ LEC's
and propose a model for the vector meson resonances which
implements flavour symmetry breaking (to first order) in a more general way.
A phenomenological determination of all the parameters entering
this resonance Lagrangian is performed and the complete contribution
of order $p^6$ in terms of the basis of ref.\cite{bce99class} is
worked out.
Tests of this modelling are performed by comparing with resonance
contributions in unsubtracted sum rules. We finally identify a combination
of LEC's which should be weakly sensitive to the scalar resonance sector
and discuss the result.

\section{Results on $C_1^r$ to $C_4^r$}

\subsection{Notation}
At first, let us recall some standard results and notation concerning
the $\pi K$ scattering amplitude.
Assuming isospin symmetry to be exact, $\pi K$ scattering is described
in terms of two independent isospin amplitudes $F^I(s,t,u)$, with
$I=1/2,\ 3/2$ and the Mandelstam variables, $s,t,u$, satisfy
\be
s+t+u=2\Sigma,\  \Sigma=\mkd+\mpid\ .
\en
Under $s,\ u$ crossing the following relation holds,
\be
F^{1\over2}(s,t,u)=-{1\over2}F^{3\over2}(s,t,u)+
                    {3\over2}F^{3\over2}(u,t,s)\ .
\en
It is then convenient to form the two combinations
$F^+$ and $F^-$ which are respectively even and odd under $s,\ u$ crossing,
\bea\lbl{fpmdef}
&&F^+(s,t,u)={1\over3} F^{1\over2}(s,t,u)
+            {2\over3} F^{3\over2}(s,t,u)\nonumber\\
&&F^-(s,t,u)={1\over3} F^{1\over2}(s,t,u)
-            {1\over3} F^{3\over2}(s,t,u)\ .
\ena
Under $s,\ t$ crossing $F^+$ and $F^-$ are simply proportional to the $I=0$
and the $I=1$ $\pi\pi\to K\Kbar$ amplitudes,
\bea\lbl{stcross}
&&G^0(t,s,u)=\sqrt6 F^+(s,t,u)\nonumber\\
&&G^1(t,s,u)=2      F^-(s,t,u)\ .
\ena
A region where one expects ChPT to apply is around the sub-threshold point
$t=0$, $s=u=\mkd+\mpid\;$. The $\pi K$ amplitude can be characterized in the
neighbourhood of this point  by performing an expansion in powers of
$t$ and $s-u$\cite{lang}.
The subthreshold coefficients $C^\pm_{ij}$ are
dimensionless quantities defined from this expansion
\bea
&& F^+(s,t,u)= \sum_{i j}\, C^+_{ij}\,
{ t^i \nu ^{2j} \over m_{\pi^+}^{2i+2j}}\ ,
\nonumber\\
&& {F^-(s,t,u)\over\nu}
= \sum_{i j}\, C^-_{ij}\, { t^i \nu ^{2j} \over m_{\pi^+}^{2i+2j+1}}\ ,
\ena
with
\be
\nu= {s-u\over 4m_K}\ .
\en
\subsection{Chiral $O(p^6)$ tree level contributions to the
sub-threshold coefficients}

The contributions at tree level from the $O(p^6)$ chiral Lagrangian
to the sub-threshold coefficients have been worked out in \cite{bijdontpik}
and can be found explicitly in this reference. We will discuss what can be
learned about the $O(p^6)$ LEC's $C^r_i$ from these expressions. Let us
begin by noting some general features of the correspondence between the
sub-threshold coefficients and the LEC's. At first, the coefficients
such that
\be
C^+_{ij}:\ i+2j \ge 4,\quad C^-_{ij}:\ i+2j \ge 3
\en
get no contribution at all from the $O(p^6)$ LEC's. This implies that the
chiral expressions at order $p^6$ for these coefficients obey
convergent unsubtracted dispersions relations. As a simple example
$C^+_{02}$ can be written as (which is easily derived from eq.\rf{fpdisp}
below),
\be\lbl{cp02disp}
\left.C^+_{02}\right\vert_{p^4+p^6} ={32 m_K^4  m_\pi^4 \over \pi}
\!\int _\mpd ^\infty ds'\,{ Im F^+(s',0)_{p^4+p^6} \over (s'-\Sigma)^5 }
\en
(with $m_+=\mk+\mpi$).
In this expression one can compute $Im F^+(s',0)_{p^4+p^6}$ by expanding over
partial waves and, for each partial-wave amplitude, using the chiral expansion
of the unitarity relation
\bea\lbl{pwunit}
&& Im f_l^I(s')_{p^4+p^6}= \frac{\sqrt\lambda}{s} f_l^I(s')_{p^2} \Big[
   f_l^I(s')_{p^2}
\nonumber\\
&& \quad +2 {\rm Re }  f_l^I(s')_{p^4} \Big]\ .
\ena
In this manner, we could reproduce precisely the numerical result
$C^+_{02}=0.23$ obtained  in ref.\cite{bijdontpik}.

Next, the chiral expressions for the set of coefficients which satisfy
\be\lbl{6deriv}
C^+_{ij}:\ i+2j=3,\quad C^-_{ij}:\ i+2j=2
\en
involve the four LEC's $C_1^r$,  $C_2^r$,  $C_3^r$,  $C_4^r$\cite{bijdontpik}
which are associated with the following four chiral Lagrangian terms
(the definitions of the various chiral building blocks $u_\mu$,
$h_{\lambda\nu}$ etc... which appear below can be found, for instance,
in ref.\cite{bce99class})
\bea\lbl{lag0chi}
&& O_1= \trace{ u_\mu\, u^\mu\, h_{\lambda\nu}\, h^{\lambda\nu} }\nonumber\\
&& O_2= \trace{u_\mu\, u^\mu}\trace{h_{\lambda\nu}\, h^{\lambda\nu}}\nonumber\\
&& O_3= \trace{h_{\mu\nu}\, u_\rho\, h^{\mu\nu}\, u^\rho }\nonumber\\
&& O_4= \trace{h_{\mu\nu}(u_\rho\, h^{\mu\rho}\, u^\nu
+u_\nu\, h^{\mu\rho}\, u_\rho)}
\ .
\ena
These terms contain six derivatives and do not involve quark masses.
We will discuss below the
determination of these LEC's obtained from the sub-threshold $\pi K$ amplitudes
as well as from $\pi \pi$ amplitudes.

We next consider the sub-threshold coefficients which satisfy
\be
C^+_{ij}:\ i+2j=2,\quad C^-_{ij}:\ i+2j=1\ ,
\en
i.e. the three coefficients $C^+_{20}$, $C^+_{01}$,
$C^-_{10}$. Their chiral
expansions involve, in addition to  $C^r_1$, $C^r_2$, $C^r_4$ the
eight LEC's $C^r_5\cdots$ $C^r_{8}$, $C^r_{10}\cdots C^r_{13}$
and the three LEC's  $C^r_{22}$,  $C^r_{23}$
$C^r_{25}$. We reproduce the corresponding Lagrangian terms below
for the convenience of the reader
\begin{align}\lbl{lag1chi}
& O_5=   \trace{(u_\mu u^\mu)^2 \Chip }\ \
&& O_6=   \trace{(u_\mu u^\mu)^2}\trace{\Chip }\
\notag\\
& O_7=   \trace{u_\mu u^\mu}\trace{u_\nu u^\nu \Chip}
&& O_8=   \trace{u_\mu u^\mu u_\nu \Chip u^\nu}\ \
\notag\\
&O_{10}= \trace{\Chip u_\mu u_\nu u^\mu u^\nu}\ \
&&O_{11}=\trace{\Chip}\trace{u_\mu u_\nu  u^\mu u^\nu}
\notag\\
&O_{12}=\trace{h_{\mu\nu}\, h^{\mu\nu}\, \Chip   }\ \
&&O_{13}=\trace{h_{\mu\nu}\, h^{\mu\nu}}\trace{\Chip}\ \
\notag\\
&O_{22}=i\trace{ \Chim\{ h_{\mu\nu},u^\mu u^\nu\} }
&&O_{23}=i\trace{ \Chim  h_{\mu\nu}}\trace{u^\mu u^\nu}\ \
\notag\\
&O_{25}=i\trace{h_{\mu\nu} u^\mu \Chim u^\nu }\ .
\end{align}
These terms contain four derivatives and
a single insertion of the quark mass matrix. The information provided by the
$\pi K$ amplitude is not sufficient to determine separately
all these LEC's. Previously,
the LEC $C_{12}$ (as well as the LEC $C_{34}$) have been determined based
on the $\Delta S=1$ scalar form-factor\cite{jopOp6} by combining the
chiral $O(p^6)$ calculations of ref.\cite{bijdscalff} with the dispersive
construction method of ref.\cite{dgl}. Constraints on  $C_{12}$
and  $C_{13}$ have also been obtained from $\Delta S=0$
scalar form-factors\cite{bijdscalff}.

Finally, the coefficients with $i+2j=0,1$: $C^+_{00}$, $C^+_{10}$,
$C^-_{00}$ involve
thirteen more $O(p^6)$ LEC's among those associated with chiral Lagrangian
terms containing two or three insertions of the quark mass matrix.

\subsection{Determination of $C_1^r \cdots C_4^r$}
The set of sub-threshold coefficients defined in eq.\rf{6deriv} constrain the
values of the four LEC's $C_1^r \cdots C_4^r$. The relevant formulas
from ref.\cite{bijdontpik} read
\bea\lbl{relform}
&& \left.C_{30}^+\right\vert_{C_i}=
{1\over2}\,(-7 C_1^r-32 C_2^r +2 C_3^r+10 C_4^r)\, {m_\pi^6\over F_\pi^4 }
\nonumber\\
&& \left.C_{11}^+\right\vert_{C_i}=
     8\,(  3 C_1^r+ 6 C_3^r         -2 C_4^r)\,{m_\pi^4 m_K^2\over F_\pi^4}
\nonumber\\
&& \left.C_{20}^-\right\vert_{C_i}=
       6\,(  - C_1^r          +2C_3^r +2 C_4^r)\,{m_\pi^5 m_K\over F_\pi^4}
\nonumber\\
&& \left.C_{01}^-\right\vert_{C_i}=
    32\,(  - C_1^r          +2C_3^r +2 C_4^r)\,{m_\pi^3 m_K^3\over F_\pi^4}
\ .
\ena
The $O(p^4)$ LEC's $L_i^r$ contribute to these coefficients only via one-loop
diagrams such that one may use for the $L_i's$ the numerical values determined
at  $O(p^4)$. The last two equations \rf{relform} involve the same combination
of LEC's $C_i^r$. Therefore, we can determine three combinations, for instance
$C_1^r+4C_3^r$, $C_2^r$, $C_4^r+3C_3^r$.

Independent informations are provided
by $\pi\pi$ scattering. The $\pi\pi$ amplitude constrains the two combinations
$C_4^r+3C_3^r$ and $C_1^r+4C_3^r+2C_2^r$. The numerical values which we quote
in table\ \Table{C1a4res} make use of the expressions from
\cite{bijcege,bijdontpipi}
\bea\lbl{r5r6}
&& r_5^r= F_\pi^2\, (-8C_1^r-16 C_2^r+10 C_3^r+14 C_4^r)
+\nonumber\\
&&{23\fpid\over15360\,\pi^2\mkd} +log's
\\
&& r_6^r= F_\pi^2\, ( 6C_3^r+2C_4^r) +{\fpid\over15360\,\pi^2\mkd} +log's
\nonumber\ena
and the numerical values for $r_5^r$, $r_6^r$ obtained in ref.\cite{cglpipi}
from a Roy equations analysis.
The right-hand sides of eqs.\rf{r5r6} involves a quadratic polynomial
in
$\log(\mkd/\mud)$ and $\log(\metad/\mud)$
which we have not determined. We have
attempted to minimize its influence by performing the matching at a scale
$\mu^2=m_K m_\eta $ before evolving the scale to $M_\rho$.

\begin{table*}
\centering
\begin{tabular}{|c|cccc|}\hline
input & $C_1^r+4C_3^r$ & $C_2^r$      &$C_4^r+3C_3^r$&$C_1^r+4C_3^r+2C_2^r$ \\
\hline
$\pi K: C^+_{30}, C^+_{11}, C^-_{20}$
      & $20.7\pm 4.9$  & $-9.2\pm 4.9$ & $9.9\pm 2.5$&     $ 2.3\pm 10.8$   \\
$\pi K: C^+_{30}, C^+_{11}, C^-_{01}$
      & $28.1\pm 4.9$  & $-7.4\pm 4.9$ & $21.0\pm 2.5$&    $13.4\pm 10.8$ \\
$\pi\pi$
      & \              &               & $23.5\pm 2.3$&    $18.8\pm 7.2$  \\
Resonance model
      & $7.2$           & $-0.5$        & $10.0$      &    $6.2$      \\ \hline
\end{tabular}
\caption{\sl Results for combinations of
$C_1^r(\mu)$ to $C_4^r(\mu)$ with $\mu=0.77$ GeV in units of
$10^{-4}\ GeV^{-2}$ derived from the $\pi K$ subthreshold parameters.  Also
shown are results based on the $\pi\pi$ amplitude and from a resonance model}
\lbltab{C1a4res}
\end{table*}

It is of interest to compare these results from those of the
resonance saturation model.
In the case of $C_1^r$,..., $C_4^r$ it suffices
to consider resonances in the chiral limit as was the case for the $O(p^4)$
LEC's \cite{egpr}. If one uses simply the same Lagrangian as in ref.\cite{egpr}
(which was also used in the $\pi K$ analysis of ref.\cite{bijdontpik})
one obtains
\begin{align}
  C_1^{V+S}&= {G_V^2\over 8M_V^4} - {c_d^2\over 4M_S^4 }\ ,
  & C_3^{V+S}=& 0 \notag\\
  C_2^{V+S}&= {c_d^2\over 12M^4_S } -{\tilde c_d^2\over 4M_{S_1}^4}\ ,
  & C_4^{V+S}=& {G_V^2\over 8M_V^4}\ .
\end{align}
Contributions from resonance Lagrangian terms like
\be\lbl{sixderiv}
\trace{ \nabla^\lambda V_{\lambda\mu}[ h_{\mu\nu},u^\nu]},\
\trace{ \nabla^\lambda V_{\mu\nu}    [ h_{\mu\lambda},u^\nu]}
\en
should, in principle, also be considered but we will not do so
here\footnote{While this paper was being completed a preprint
appeared\cite{kaiser} containing a general discussion of resonance
Lagrangian terms contributing at order $p^6$.}.
In discussing such higher
derivative terms, it is important to implement proper asymptotic conditions.

Numerical values are shown in table 1, using the same values for the couplings
as in ref.\cite{egpr}, i.e.
\be
G_V=53\, \MeV,\  c_d=32\, \MeV,\  c_m=42\, \MeV.
\en
We note
that this value of $G_V$ is somewhat smaller than the one which derives
from the $\rho\to 2\pi$ width ($G_V\sim 64.1$ MeV, see sec.3 ) but was
shown to yield good results for the $O(p^4)$ LEC's.
In the case of $C_2$,
which is OZI suppressed we show, for illustration, the value
derived from the OZI violation model A of ref.\cite{cirigscal}.
The table also shows that the results obtained using $C_{20}^-$ as input
and those using $C_{01}^-$ are compatible for  $C^r_1+4 C^r_3$ and
for $C^r_2$ but not quite so for $C^r_4+3 C^r_3$. The error, however,
does not take higher order chiral effects into account. The results which use
$C_{01}^-$ are compatible with the $\pi\pi$ results.
The simplest resonance saturation model is seen to give  correct signs
and order of magnitudes for the LEC's shown in table 1 but the agreement
is certainly not as good as in the case of the $O(p^4)$ couplings.

\section{Symmetry breaking in the vector meson chiral Lagrangian revisited}

\subsection{Observation of some discrepancies}
Let us now turn our attention to the three coefficients $C_{20}^+$,
$C_{01}^+$, and $C^-_{10}$. As mentioned above, their chiral expansions
get tree level contributions from the Lagrangian terms,
$O_{5}$,...,$O_{13}$ and $O_{22}$,...,$O_{25}$ which contain
four derivatives and one quark mass factor. Their chiral expansions also
receive $O(p^4)$ tree level contributions involving the LEC's $L_1$, $L_2$
$L_3$. In general, in such a situation,
the hope is that one may use a resonance model
estimate for the $O(p^6)$ LEC's  and then derive improved determinations
for the  LEC's $L_i$. This idea was actually followed in the series of papers
\cite{ABT2pt,ABTKl4a,ABTKl4b} which used as experimental input the
pseudo\-scalar meson masses, decay constants and the $K_{l4}$
decay form-factors.
Using the determination of  the chiral coupling constants
obtained in these references from this procedure, the three  $\pi K$
sub-threshold coefficients can be predicted.
The results obtained in \cite{bijdontpik} are reproduced in
table \Table{3coeffs}.  Looking at table \Table{3coeffs} it is rather
striking that there is a serious discrepancy,
for all these three sub-threshold
coefficients, between the chiral predictions and the dispersive calculations.

\begin{table}[hb]
\centering
\begin{tabular}{|c|cccc|}\hline
\ & $(p^4)_{L_i=0}\!$ & $(p^6)_{L_i=C_i=0}\!$ & $(p^4+p^6)_{total}\!\!\!\!$
& Dispersive \\ \hline
$C_{20}^+$ & 0.0255 & -0.0254 & 0.003 & $0.024\pm 0.006$ \\
$C_{01}^+$ & 1.673  &  1.492  & 3.8   & $2.07\pm  0.10 $ \\
$C_{10}^-$ &-0.0253 &  0.121  & 0.09  & $0.31\pm 0.01$ \\ \hline
\end{tabular}
\caption{\sl Comparison of the dispersive results for three sub-threshold
parameters (last column) with the chiral calculation of ref.\cite{bijdontpik}
at order $p^6$. The second and third columns display  results
obtained when the LEC' $L_i^r(\mu)$ and $C_i^r(\mu)$
are set equal to zero at $\mu=0.77$ GeV. The fourth column displays the full
chiral result from ref.\cite{bijdontpik}
}
\lbltab{3coeffs}
\end{table}
\subsection{Should one blame the dispersive representations ? }
A possible explanation for these discrepancies
could be that the dispersive calculations are not correct. Let us argue,
considering the  particular example of $C^+_{01}$
which is rather simple,
that this is unlikely to be the case.
One may start with a fixed$-t$ dispersive representation, at $t=0$,
of the amplitude $F^+(s,t)$ with two subtractions,
\bea\lbl{fpdisp}
&& F^+(s,0)= c^+(0) +{1\over\pi}\int_\mpd^\infty ds' \Big[{1\over s'-s}
+{1\over s'-u} \nonumber\\
&&\quad -{ 2(s'-\Sigma)\over (s'-\mpd)(s'-\mmd)} \Big]
Im F^+(s',0).
\ena
The validity of this kind of dispersion relation as well as that of the
Froissart bound which ensures convergence can be established in a rigorous
manner\cite{dispvalid,froissart}. From eq.\rf{fpdisp} it is straightforward
to derive the following sum rule for the sub-threshold parameter  $C^+_{01}$
\be\lbl{cp01sum}
C^+_{01}={8\mkd \mpid\over\pi}\int_\mpd^\infty {Im F^+(s',0)\over
(s'-\Sigma)^3 }\,ds'\ .
\en
(Note that unlike the case of $C^+_{02}$, this sum rule is useless for
deriving the chiral result.) The integrand needed in this sum rule is
displayed in fig. \fig{cp01int}.
\begin{figure*}
\centering
\includegraphics[width=12cm]{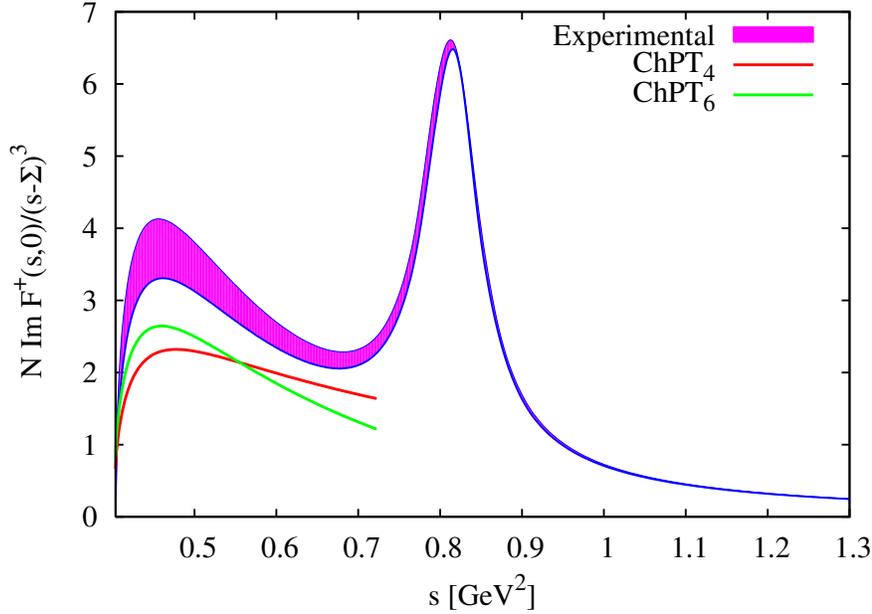}
\caption{\sl Integrand to be used in the sum rule, eq.\rf{cp01sum} }
\lblfig{cp01int}
\end{figure*}
The following remarks can be made.
The contributions from the high energy region $\sqrt{s'}\ge 2 $ GeV
are negligibly small.
Most of the contributions are from the $S$ and
the $P$ waves, they are concentrated in the region
$\sqrt{s'}\lapprox 1 $ GeV and there
are no numerical difficulties in computing the integral.
The $S$ wave in the lower energy range is the part affected with
the largest error. In this region, one can compare with ChPT calculations
(which up to order six do not depend on the LEC's $C_i^r$): the
difference is of the order of 20\% at most and the ChPT result for
$Im F^+(s',0)$ tends to be smaller and not larger than the one derived
from experiment.
In conclusion, this sum rule seems fairly solid: any reasonable fit
to the experimental data of refs.\cite{estabrooks,aston} will give a number
$C^+_{01}\simeq 2$.

\subsection{Vector resonance chiral Lagrangian}
Another possible cause for the discrepancies revealed in table \Table{3coeffs}
could be that the resonance saturation model used to evaluate
the LEC's $C_i$ needs to be improved.
The coefficients in table \Table{3coeffs}
are associated with  $O(p^6)$ operators containing
one quark mass factor such that the
corresponding LEC's are sensitive to flavour symmetry breaking of the
light resonances. We will re-examine the case of the vector mesons here,
and follow the approach to first construct a Lagrangian containing the
resonance fields and then integrate them out. A convenient method for this
construction (see e.g. \cite{georgibook}) is to make use of  non-linear
representations of the chiral group\cite{cwz}. For the purpose of
generating chiral Lagrangian terms it is also convenient to adopt a homogeneous
chiral transformation rule for all the resonances
\be
R\to h[\phi]\, R\, h[\phi]^\dagger\ .
\en
Such a transformation rule ensures that one can ascribe a definite
chiral order to each resonance field. A detailed discussion
in connection with the $O(p^4)$ LEC's can be found in refs.\cite{egpr,eglpr}.
In ref.\cite{bijdontpik} the vector field formalism (see \cite{eglpr}) was
used and flavour symmetry breaking is described via a single term
\be
{\cal L}^m_{V}= f_\chi\trace{ V_\mu [u^\mu,\Chim] }\ .
\en
This term is the unique one relevant to the $O(p^6)$ LEC's because
in the vector field formalism the field $V_\mu$ has chiral order three. The
coupling constant $f_\chi$ was determined such as to reproduce the experimental
value for the ratio $\Gamma( K^* \to K\pi)/ \Gamma(\rho\to \pi\pi)$
which gave\lbl{bijsain}
\be
f_\chi= -0.025\ .
\en
In seeking
for an improvement we note that, in this formalism,
the symmetry breaking effects induced from the masses of the vector mesons
are absent at order $p^6$, which seem somewhat unnatural. This suggests
to investigate different formalisms. A discussion of symmetry
breaking based on the  massive Yang-Mills approach was performed in
ref.\cite{schechter}.  We will make use here of the formalism
which uses anti-symmetric
tensors\cite{gl84,egpr} instead of vector fields\footnote{In ref.\cite{eglpr}
it was shown how these two formalisms can be
made to give exactly equivalent results for the $O(p^4)$ LEC's. This
necessitates that a number of asymptotic constraints for Green's functions,
form-factors or scattering amplitudes be implemented.}.
The part which is
relevant for the $O(p^4)$ LEC's was considered in ref.\cite{egpr}
\bea\lbl{lag0}
&&{\cal L}_{AT}^0= -{1\over2}\trace{ \nabla^\lambda V_{\lambda\mu}
\nabla_\nu V^{\nu\mu} }+{1\over 4} \mvd \trace{ V_{\mu\nu} V^{\mu\nu} }
\nonumber\\
&&\phantom{ {\cal L}^0= }
 +{F_V\over2\sqrt2}\trace{ V_{\mu\nu} f^{\mu\nu}_+ }
 +{i\,G_V\over\sqrt2}\trace{V_{\mu\nu} u^\mu u^\nu}\ .
\ena
From eq.\rf{lag0} one can deduce the chiral order of the resonance field
\be\lbl{order}
V_{\mu\nu}\sim O(p^2)\ .
\en
As a consequence, the kinetic energy term in eq.\rf{lag0} is $O(p^6)$
while the other terms are $O(p^4)$. Let us now consider all possible
terms which are chiral symmetry breaking corrections to the terms in
eq.\rf{lag0}. Neglecting OZI rule violation, we find that there are six
independent such terms which have chiral order six (some of these have
been considered also recently in ref.\cite{leupold})
\bea\lbl{efflag}
&&{\cal L}^m_{AT} =   \\
&&\quad {1\over2}e_V^m\trace{\Chip V^{\mu\nu} V_{\mu\nu}}
+i{g^m_{V1}\over \mv } \trace{ V^{\mu\nu} \{\Chip , u_\mu u_\nu \} }
\nonumber\\
&&\quad +i{g^m_{V2}\over \mv } \trace{ V^{\mu\nu} u_\mu\Chip u_\nu}
-{f_\chi\over \mv} \trace{ \nabla_\mu V^{\mu\nu} [\Chim,u_\nu]}
\nonumber\\
&&\quad +{f^m_{V1}\over \mv} \trace{ V_{\mu\nu} \{f^{\mu\nu}_+, \Chip\} }
+{f^m_{V2}\over \mv} \trace{ V_{\mu\nu} [f^{\mu\nu}_-, \Chim] }\ .
\nonumber\ena
Only the first four of these terms play a role in $\pi K$ scattering.
Instead of a single coupling constant, $f_\chi$, in the vector formalism,
one has four independent couplings here: $e^m_V$, $g^m_{V1}$, $g^m_{V2}$,
$f_\chi$. Let us now discuss the determination of these couplings
from experiment.

\subsection{Determination of the vector Lagrangian coupling constants}

\subsubsection{Determination of $e_V^m$}
At first, it is not difficult to determine $e_V^m$ based on the mass relations
\bea
&& M^2_\rho = M^2_V +8 e_V^m B_0 \hat m\nonumber\\
&& M^2_{K^*}= M^2_V +4 e_V^m B_0 (m_s + \hat m)\ .
\ena
Isospin breaking is neglected here and we have denoted
$\hat m = (m_u+m_d)/2$. For the numerics,
we can use the results from the chiral expansion
at leading order,
\bea
&& 2B_0\mhat = M^2_{\pi^0}= (134.98\,\MeV)^2\nonumber\\
&& {m_s\over \mhat} = {M^2_{K^+}+M^2_{K^0}-M^2_{\pi^+}\over M^2_{\pi^0}}
\simeq 25.90 \nonumber\\
&& B_0 (m_u-m_d)=M^2_{K^+}-M^2_{K^0}-M^2_{\pi^+}+M^2_{\pi^0}\nonumber\\
&& \phantom{B_0 (m_u-m_d)}
\simeq -0.285\, M^2_{\pi^0}\ .
\ena
Using the experimental values of the $K^*(892)$ and the $\rho(770)$ masses
we obtain
\be
e_V^m\simeq 0.22\ .
\en
If, instead, one uses the masses of the $\phi(1020)$ and the $\rho(770)$
mesons one would obtain $e_V^m\simeq 0.24$, suggesting that the error
should ne reasonably small for this quantity.

\subsubsection{Determination of $f_\chi$ and $\fvdeux$}
As a next step we consider the coupling constant
$f_\chi$. In ref.\cite{bijdontpik} $f_\chi$ was related to symmetry
breaking in the decays of vectors into two pseudo\-scalars. Here, we will
argue that these decays determine the two couplings $g^m_{V1}$, $g^m_{V2}$.
Concerning $f_\chi$, a physically plausible estimate can be obtained
by relating it to the
decay of the $\pi(1300)$ resonance.
Let us denote the $\pi(1300)$ nonet matrix  by $P$ and consider the
Lagrangian,
\bea
&&{\cal L}^{\pi(1300)}= {1\over2}\trace{\nabla^\mu P\nabla_\mu P }
-{1\over2} M_P^2\trace{P^2}
 +i d_m \trace{P\Chim} +
\nonumber\\
&&\quad iG'_V   \trace{\nabla_\mu V^{\mu\nu} [ P, u_\nu] }
 +iG''_V \trace{           V_{\mu\nu} [f_-^{\mu\nu}, P]}\ .
\ena
This extends the Lagrangian considered in ref.\cite{egpr} by the last
two terms proportional to $G'_V$ and $G''_V$ respectively and
which have chiral order equal to six.
Integrating out the $\pi(1300)$ meson,
one finds that the couplings $f_\chi$ and
$f^m_{V2}$ which were appearing in eq.\rf{efflag} are proportional
respectively to $G'_V$ and $G''_V$
\be
f_\chi =  G'_V {d_m M_V\over M^2_P },\quad
f^m_{V2} = G''_V  {d_m M_V\over M^2_P }\ .
\en
The coupling $d_m$ was introduced in ref.\cite{egpr}. It can be estimated
by appealing to a chiral super-convergence
sum-rule associated  with the correlator of two scalar currents minus
the correlator of two pseudo\-scalar currents (see ref.\cite{gl84} ).
Saturating the sum rule from the contributions of the pion, the $\pi(1300)$ and
the $a_0(980)$ one gets the relation
\be
8d_m^2+ F_0^2 -8c_m^2 =0\ .
\en
Using the value  $c_m\simeq 42$ MeV which was obtained ref.\cite{egpr}
then gives
\be
d_m\simeq 26 \ {\rm MeV}\ .
\en
The coupling $G'_V$  can be related  to  the decay amplitude
$\pi(1300)$ $\to \rho\pi$,
\be 
\Gamma_{\pi(1300)\to \rho\pi}= { 2(G'_V)^2 p_{CM}^3\over \pi F^2_\pi }\ .
\en
The total width of the $\pi(1300)$ is known to be
rather large but has not actually been very precisely
determined (the PDG\cite{pdg04} quotes a range of values between 200 and
600 MeV).
For definiteness, let us use the result obtained in ref.\cite{abele}
who also find the $\rho\pi$ decay mode to be the dominant one
\be
\Gamma_{\pi(1300)\to \rho\pi}\simeq 268\pm 50 \ {\rm MeV}\ .
\en
This gives the estimate
\be
\vert G'_V\vert \simeq 0.23
\en
yielding
\be\lbl{fchinum}
\vert f_\chi \vert\simeq  2.8\,10^{-3}\ .
\en
This value is one order of magnitude
smaller than the one obtained in ref.\cite{bijdontpik}.
One consequence concerns the lifetime of the $\pi K$ atom
which receives a contribution (via resonance saturation of the
LEC's $C_i$ ) which is quadratic in $f_\chi$.
If one uses the numerical value \rf{fchinum} for $f_\chi$,
the size of the $O(p^6)$ contribution to the lifetime
is rather small (see the detailed
discussion in ref.\cite{julia}).

A somewhat different
approach is to consider the 3-point correlation function $<VAP>$, model it
in terms of a finite number of resonances, and
constrain the coupling constants in order to enforce the proper QCD asymptotic
conditions\cite{mouss97,knnyff}. This was reconsidered
recently by Cirigliano et al.\cite{ciriglvap} who improved on earlier work
by including the $\pi(1300)$
nonet in the construction together with the vector, axial-vector and pion
multiplets. In this manner,
they have obtained  a determination of the $\pi(1300)$ couplings
$G'_V$, $G''_V$ in terms of the vector and axial-vector
resonance masses,
\be
G'_V=- {\sqrt{M^2_A -M^2_V}\over 2M_A},\quad
G''_V=- {\sqrt{M^2_A -M^2_V}\over 8M_A}\ .
\en
Using $M_A= \sqrt{2} M_V$ this gives
\be
f_\chi\simeq  -4.2\,10^{-3},\quad f^m_{V2}\simeq -1.1\,10^{-3}\ .
\en
This method provides a determination of $f_\chi$ which
is in reasonably good agreement
with the one based on the $\pi(1300)$ decay width and gives also
the sign as well as a determination of the coupling $f^m_{V2}$.

\subsubsection{Determination of $g^m_{V1}$}
The coupling $g^m_{V1}$ can be determined from the decay amplitudes of
vector mesons into two pseudo\-scalars. The decay amplitudes have the
following form
\be
{\cal T}(V_a\to\phi_b\,\phi_c)= M_{V_a} \epsilon\cdot(p_1-p_2)\, T_{abc}\ .
\en
Correspondingly,  the decay width is given by
\be
\Gamma(V_a\to\phi_b\,\phi_c)= \vert T_{abc}\vert^2\, {p_{cm}^3\over 6\pi}\ .
\en
Using the Lagrangian \rf{efflag} these amplitudes get expressed as a
function of two combinations of the couplings
$g^m_{V1}$, $g^m_{V2}$ and $f_\chi$ for which we introduce the notation
\bea
&&\ghvun=  g^m_{V1} +{1\over2} f_\chi
\nonumber\\
&&\ghvdeux=g^m_{V2} +          f_\chi\ .
\ena
The amplitude for $\rho^+\to \pi^+\pi^0$, at first, reads
\bea
&& T_{\rho^+\to \pi^+\pi^0}= {1\over F^2_\pi}\, G_V^{eff}\ ,\quad
\nonumber\\
&& G_V^{eff}=G_V + {4\sqrt2\mhat B_0\over M_V} (2\ghvun+\ghvdeux) \ .
\ena
Using the experimental values  for the mass $m_\rho= 775.5\pm0.5 \, \MeV$
and the width $\Gamma=150.2\pm2.4\, \MeV$ from ref.\cite{pdg04} gives
\be
G_V^{eff}\simeq 65.8 \ {\rm MeV}\ .
\en
Next, we consider the decays $K^*\to K\pi$ and $\phi\to K\bar{K}$.
\bea\lbl{Vdecays}
&&T_{K^{*+}\to K^0\pi^+}= \nonumber\\
&&\quad {\sqrt2\over 2 F_{K^0}F_{\pi^+}}\Big\{
G_V^{eff}+{4\sqrt2\ghvun\over M_V} B_0 (\ms-\mhat)\nonumber\\
&&\quad +{2\sqrt2(\ghvun-\ghvdeux)\over M_V} B_0\mdiff \Big\}
\nonumber\\
&&T_{K^{*+}\to K^+\pi^0}= \nonumber\\
&&\quad{1\over 2 F_{K^+}F_{\pi^0}}\Big\{G_V^{eff}
+{4\sqrt2\ghvun\over M_V} B_0 (\ms-\mhat) \nonumber\\
&&\quad+{2\sqrt2(\ghvun+\ghvdeux)\over M_V} B_0\mdiff \Big\}
\nonumber\\
&&T_{\phi\to K^+ K^-}=
-{\sqrt2\, e^2 F_V\over 6M^2_\phi}{2M^2_\rho-M^2_\phi\over M^2_\rho-M^2_\phi}
\nonumber\\
&&\quad +{\sqrt2\over 2 F_{K^+}^2 } \Big\{G_V^{eff}
+{8\sqrt2\ghvun \over M_V} B_0 (\ms-\mhat)\nonumber\\
&&\quad +{2\sqrt2\ghvdeux\over M_V} B_0\mdiff\Big\}
\nonumber\\
&&T_{\phi\to K^0 {\Kbar^0}}=
{\sqrt2\, e^2 F_V\over 6M^2_\phi}{ M^2_\phi\over M^2_\rho-M^2_\phi}\nonumber\\
&&\quad +{\sqrt2\over 2 F_{K^0}^2 } \Big\{G_V^{eff}
+{8\sqrt2\ghvun \over M_V} B_0 (\ms-\mhat)\nonumber\\
&&\quad -{2\sqrt2\ghvdeux\over M_V} B_0\mdiff \Big\}\ .
\ena
These expressions include  isospin breaking contributions proportional
to $m_u-m_d$ and those proportional to $e^2 F_V$ induced by the coupling of the
neutral vector mesons to the photon. We have also taken into account
the influence of wave-function renormalization of the pseudo\-scalar mesons.
If we ignore iso\-spin breaking, i.e. set $m_u=m_d$, then the  decay amplitudes
\rf{Vdecays} no longer depend on $\ghvdeux$ which allows us to determine
$\ghvun$. Combining the experimental values\cite{pdg04}
for the $K^{*+}$ and $K^{*0}$ decay
widths into $K\pi$ we obtain
\be
\ghvun\simeq 6.0\,10^{-3}.
\en
If one uses the $\phi$ decay widths into $K^+ K^-$ and $K^0\bar{K}^0$ instead,
one obtains a smaller but not very different value,
\be
\ghvun\simeq 4.3\,10^{-3}.
\en
From these two results one can infer $\ghvun = (5.2\pm 1.5)\,10^{-3} $.

\subsubsection{Determination of $g_{V2}^m$}
Finally, we have to determine $g_{V2}^m$. The results of the previous
subsection shows that if one forms isospin breaking combinations
\bea
&&T_{K^{*+}\to K^0\pi^+}-\sqrt2 T_{K^{*+}\to K^+\pi^0},\
\nonumber\\
&&T_{\phi\to K^+ K^-}-T_{\phi\to K^0 {\bar K^0}}\ ,
\ena
the coupling $g_{V2}^m$ is the only one which contributes.
In practice, however, it turns out not to be possible to determine $g^m_{V2}$
in this way.  Precise experimental information exists for isospin violation
in $\phi$ decays but, in this case, there are significant electromagnetic
contributions as well, which are difficult to evaluate.
Further amplitudes which vanish in the isospin
limit are $\omega\to\pi^+\pi^-$ and $\rho^+\to\pi^+\eta $. These amplitudes
have the following expressions,
\bea\lbl{omega}
&&T_{\omega\to\pi^+\pi^-}={G_V\over F_\pi^2 (M^2_\omega-M^2_\rho)}\times
\nonumber\\
&& \quad
\left\{ {m_u-m_d\over m_s-\mhat } (M_{K^*}^2-M_\rho^2) +{e^2 F_V^2\over3}
\right\}
\nonumber\\
&&\quad +{2\sqrt2\over F_\pi^2} {(2\ghvun-\ghvdeux)\over M_V} B_0\mdiff
\nonumber\\
&&T_{\rho^+\to\pi^+\eta}= {\sqrt3\, G_V^{eff}\over 4 F_\pi F_\eta}
{ (m_u-m_d)\over (m_s-\mhat)}
\nonumber\\
&&\quad +{2\sqrt2\over\sqrt3 F_\pi F_\eta } {\ghvdeux\over M_V} B_0\mdiff\ .
\ena
In these cases the contribution proportional to $g_{V2}^m$ can be estimated to
be relatively small such that it is again difficult to precisely extract
its value.

The coupling $\ghvdeux$ appears in the amplitude $\rho\to K\Kbar $, as
one can see from the expression,
\bea
&& T(\rho^+\to K^+\Kbar^0)= \\
&&\quad {1\over \sqrt2 F^2_K}\left\{ G_V^{eff} +
{4\sqrt2 \ghvdeux \over M_V} B_0(m_s-\mhat) \right\}\ .
\nonumber
\ena
From an experimental
point of view, one can hope to determine this amplitude from the $\tau$ decay
process $\tau\to K\Kbar \nu_\tau$.  It is customary to approximate the
dynamics of $\tau$ hadronic decays as proceeding via a
few resonances\cite{taureson}. In the case of the $K\Kbar$ channel, the
$\rho(770)$ and the $\rho(1450)$ resonances  can contribute\cite{finkmirk}.
The resonance $\rho(1450)$ has a rather small coupling to $K\Kbar$ \cite{pdg04}
and its contribution is also suppressed by phase-space
such that it seems a plausible approximation to saturate
the integrated $\tau\to K\Kbar \nu_\tau$ decay
width from just the $\rho$ contribution. In order to compute this decay
width from our resonance model we first introduce the charged vector
current matrix element
which, in the isospin limit, involves a single form-factor
\bea
&&\trace{ K^-(p_1)K^0(p_2)\vert \bar d\gamma^\mu u\vert 0}=
(p_1 - p_2)^\mu  F_V^K(s),
\nonumber\\
&&\quad  s= (p_1+p_2)^2\ .
\ena
Computing the form factor from our effective Lagrangian, we obtain
\bea\lbl{modelff}
&&  F_V^K(s)= 1+ \\
&&  \quad {F_V\over F_K^2}\left(
G_V^{eff} + {4\sqrt2\ghvdeux \over M_V}
B_0(\ms-\mhat)  \right)\,{s\over \mvd -s }\ .
\nonumber
\ena
The $\tau$ decay rate into $K\Kbar\nu_\tau $ has the following expression
\bea
&&\Gamma_{K\Kbar}= V_{ud}^2\, {G_F^2 M_\tau^5\over 768 \pi^3}
\int_{4\mkd}^\mtaud
{ds\over \mtaud }\left(1-{4\mkd\over s}\right)^{3\over2}\times
\nonumber\\
&&\quad \left(1-{s\over\mtaud}\right)^2  \left( 1+{2s\over\mtaud} \right)
\left\vert F_V^K(s)\right\vert^2\ .
\ena
In practice, the formula \rf{modelff}, which is obtained from a tree level
calculation, does not account for the $\rho$ meson width. On may
account for this effect in a phenomenological way by replacing $M_V^2$
in the propagator in eq.\rf{modelff} by $M_V^2-iM_V \Gamma(s)$.
In the energy range
relevant for $\tau$ decay we retain the contributions to the $\rho$ width
arising from the $\pi\pi$ and the $K\Kbar$ channels as well as the
$4\pi$ channel simply approximated as $\omega\pi$ which gives, in the region
$s \ge 4 \mkd $
\bea\lbl{rowidth}
&& M_V \Gamma(s) = \\
&&\quad {M_V^2\Gamma_V\over \sqrt{s}}\left[
          \left({s-4\mpid\over\mvd-4\mpid}\right)^{3\over2}
+{1\over2}\left({s-4\mkd \over\mvd-4\mpid}\right)^{3\over2}\right]
\nonumber\\
&&\quad +{G_{\omega\rho\pi}^2\over4\pi} { \left[
(s-(M_\omega+\mpi)^2)(s-(M_\omega-\mpi)^2)\right]^{3\over2}
\over 24 s}\nonumber
\ena
The coupling constant $G_{\omega\rho\pi}$ may be estimated using vector
meson dominance and the experimental value of the $\omega\to\gamma\pi$
width\cite{pilkuhn}
\be
{G_{\omega\rho\pi}^2\over4\pi}\simeq 24\ {\rm GeV^{-2} }\ .
\en
Using the expression \rf{rowidth} for the imaginary part of the $\rho$
meson propagator and the experimental value\cite{pdg04}
of the $\tau\to K\Kbar\nu$ decay rate $R= (15.4\pm1.6)\,10^{-3} $ we obtain
\be\lbl{gv2}
\ghvdeux\simeq 0.015\ .
\en
Ignoring completely the $\rho$ width gives a larger value
$\ghvdeux\simeq 0.022$.
Alternatively, one may estimate $\ghvdeux$ by making use of an asymptotic
constraint, namely imposing that the form-factor
$F_V^K(s)$ goes as $1/s$ asymptotically. This yields a somewhat smaller value
$\ghvdeux\simeq 0.011$.  This discussion allows us to estimate that the
error on the estimate \rf{gv2} should be of the order of 50\%
i.e. $\ghvdeux= 0.015\pm 0.007$.

\subsubsection{Determination of $\fvun$ }
Finally, let us consider $\fvun$. This parameter controls flavour symmetry
breaking in the matrix elements of the vector current between a vector meson
and the vacuum,
\be
F_{K^*}-F_\rho = {8\sqrt2\, \fvun \over M_V} B_0 (m_s-\mhat)\ .
\en
We can extract the relevant information from the $\tau$ decay processes
$\tau\to \rho^- \nu_\tau$ and $\tau \to K^{*-} \nu_\tau$. Using the
experimental results from \cite{pdg04} we obtain
\be\lbl{froexp}
F_\rho = 146.3 \pm 1.2 \ {\rm MeV},\quad
F_{K^*}= 155.1 \pm 4.0 \ {\rm MeV}\ ,
\en
from which we finally deduce
\be
\fvun=0.0027\pm 0.0013\ .
\en
\subsection{Vector meson contributions to the LEC's }
Let us now integrate out the vector meson from the Lagrangian
\rf{lag0},\rf{efflag}
and consider the $O(p^6)$ chiral Lagrangian terms which are generated. One
finds
\begin{align}\lbl{chirv}
\lag_{AT}^{(6)}&= {\gvd\over4 \mvq}\trace{\nabla_\lambda [u^\lambda,u^\mu]
\nabla_\nu [u^\nu,u_\mu] }\notag\\
&-\left( {\evm\gvd\over2\mvq}-{\sqrt2G_V\gvun\over M_V^3}\right)
\trace{[u_\mu,u_\nu] u^\mu u^\nu \Chip}\notag\\
&+ \frac{G_V \gvdeux}{\sqrt{2}M_V^3} \trace{[u_\mu,\, u_\nu] u^\mu
\Chip u^\nu} \notag\\
&- \frac{G_V f_\chi}{\sqrt 2 M_V^3}i\trace{
\nabla_\mu[\Chim,\,u_\nu][u^\mu,\,u^\nu]}\notag\\
&-\frac{G_V F_V}{2 M_V^4} i \trace{\nabla_\nu [u^\nu,u^\mu]\nabla^\lambda
f_{+\lambda\mu}}
\notag\\
&- \frac{F_V^2}{4 M_V^4} \trace{\nabla^\lambda f_{+\lambda\mu}\nabla_\nu
f_+^{\nu\mu}}\notag\\
&-\frac{F_V f_\chi}{\sqrt{2}M_V^3} \trace{f_{+\mu\nu}\nabla^\mu[\Chim,u^\nu]}
\notag\\
&+\left(\frac{F_V G_V \evm}{2 M_V^4}
-\frac{2 G_V \fvun}{\sqrt{2}M_V^3}-\frac{F_V \gvun}{\sqrt{2}M_V^3}\right)\times
\notag\\
& i\trace{f_{+\mu\nu}\{\chi_+,u^\mu u^\nu\}}
-\frac{F_V \gvdeux}{\sqrt{2}M_V^3} i \trace{f_{+\mu\nu}u^\mu \chi_+ u^\nu}
\notag\\
&+\left(\frac{\evm F_V^2}{4 M_V^4} - \frac{2 F_V \fvun}{\sqrt{2} M_V^3}\right)
\trace{\Chip f_{+\mu\nu} f_+^{\mu\nu}}
\notag\\
&-\frac{F_V \fvdeux}{\sqrt{2} M_V^3} \trace{f_{+\mu\nu}[f_-^{\mu\nu},\chi_-]}
\notag\\
&-\frac{2 G_V \fvdeux}{\sqrt{2} M_V^3} i \trace{f_{-\mu\nu}[\chi_-,u^\mu u^\nu]}\ .
\end{align}
In the vector field formalism one term, proportional to  $f_\chi^2$, is
generated which does not appear in eq.\rf{chirv}. In the spirit of
ref.\cite{eglpr} we may simply add this term here\footnote{Alternatively,
one may describe spin one resonances in terms of a {\sl pair} of fields
$V_{\mu\nu}$ and $V_\mu$. A more detailed discussion of this framework
will be presented elsewhere\cite{novotkampf}.}
\be
\lag_{V}^{(6)}=-{f_\chi^2\over 2\mvd}
        \trace{[u_\mu,\Chim] [u^\mu,\Chim]}\ .
\en
In this way, we recover exactly the results of ref.\cite{bijdontpik} if
we set the extra coupling constants in our vector Lagrangian equal to zero.
Next, we can expand the chiral Lagrangian terms over the canonical
$O(p^6)$ basis established in ref.\cite{bce99class}.
After some calculation, we obtain
contributions to 45 different LEC's
\begin{align}\lbl{civ}
C_1^V &= \frac{G_V^2}{8 M_V^4},
\notag \\C_4^V &= \frac{G_V^2}{8 M_V^4},\notag\\
C_5^V &= - \frac{G_V g_{V2}^m}{\sqrt{2} M_V^3},
\notag \\C_8^V &=  \frac{\evm G_V^2}{2 M_V^4} - \frac{\sqrt{2} G_V  g_{V1}^m}{M_V^3},\notag\\
C_{10}^V &= -\frac{\evm G_V^2}{2 M_V^4} + \frac{\sqrt{2} G_V g_{V1}^m}{M_V^3} + \frac{G_V g_{V2}^m}{\sqrt{2}M_V^3},
\notag \\C_{22}^V & = \frac{G_V^2}{16 M_V^4} + \frac{G_V f_\chi}{2\sqrt 2 M_V^3},\notag\\
C_{24}^V &= \frac{1}{n}\frac{G_V^2}{4 M_V^4},
\notag \\C_{25}^V &= -\frac{3 G_V^2}{8 M_V^4}-\frac{G_V f_\chi}{\sqrt{2}M_V^3},\notag\\
C_{26}^V &= \frac{G_V^2}{4 M_V^4}-\frac{1}{n^2}\frac{G_V^2}{2M_V^4}+\frac{G_V f_\chi}{\sqrt{2}M_V^3} + \frac{f_\chi^2}{M_V^2},
\notag \\C_{27}^V &= -\frac{1}{n}\frac{G_V^2}{4M_V^4}+\frac{1}{n^2}\frac{G_V^2}{2 M_V^4},\notag\\
C_{28}^V &= \frac{1}{n^2}\frac{G_V^2}{8 M_V^4},
\notag \\C_{29}^V &= -\frac{G_V^2}{8 M_V^4}-\frac{1}{n^2}\frac{G_V^2}{4 M_V^4}-\frac{G_V f_\chi}{\sqrt{2} M_V^3}-\frac{f_\chi^2}{M_V^2},\notag\\
C_{30}^V &= \frac{1}{n^2}\frac{G_V^2}{4 M_V^4},
\notag \\C_{40}^V &= -\frac{G_V^2}{8 M_V^4},\notag\\
C_{42}^V &= -\frac{G_V^2}{8 M_V^4},
\notag \\C_{44}^V &= \frac{G_V^2}{4 M_V^4},\notag\\
C_{48}^V &= -\frac{G_V^2}{8 M_V^4},
\notag \\C_{50}^V &=  \frac{G_V F_V}{4 M_V^4}+ \frac{f_\chi F_V}{\sqrt2 M_V^3},\notag\\
C_{51}^V &= -\frac{G_V^2}{4 M_V^4}+ \frac{G_V F_V}{4 M_V^4} +\frac{f_\chi F_V}{\sqrt 2 M_V^3},
\notag \\C_{52}^V &=  - \frac{G_V F_V}{4 M_V^4}-\frac{f_\chi F_V}{\sqrt2 M_V^3},\notag\\
C_{53}^V &= - \frac{G_V F_V}{8 M_V^4} -\frac{3 F_V^2}{16 M_V^4} - \frac{f_\chi F_V}{2 \sqrt2 M_V^3},
\notag \\C_{55}^V &= \frac{G_V F_V}{8 M_V^4} + \frac{3 F_V^2}{16 M_V^4} + \frac{f_\chi F_V}{2\sqrt 2 M_V^3},\notag\\
C_{56}^V &= - \frac{G_V F_V}{4 M_V^4}+ \frac{3 F_V^2}{8 M_V^4} - \frac{f_\chi F_V}{\sqrt2 M_V^3},
\notag \\C_{57}^V &= \frac{G_V F_V}{2 M_V^4}+\frac{F_V^2}{8 M_V^4} + \frac{\sqrt2 f_\chi F_V}{M_V^3},\notag\\
C_{59}^V &= - \frac{G_V F_V}{8 M_V^4} -\frac{F_V^2}{4 M_V^4} - \frac{f_\chi F_V}{2 \sqrt2 M_V^3},
\notag \\C_{61}^V &= \frac{\evm F_V^2}{4 M_V^4} - \frac{\sqrt2 F_V \fvun}{M_V^3},\notag\\
C_{63}^V &= -\frac{\sqrt2 \fvun G_V}{M_V^3} + \frac{\evm F_V G_V}{2 M_V^4} - \frac{F_V \gvun}{\sqrt2 M_V^3},
\notag \\C_{65}^V &= -\frac{F_V \gvdeux}{\sqrt2 M_V^3},\notag\\
C_{66}^V &= \frac{G_V^2}{8 M_V^4},
\notag \\C_{69}^V &= - \frac{G_V^2}{8 M_V^4},\notag\\
C_{70}^V &= -\frac{G_V^2}{8 M_V^4}- \frac{G_V F_V}{8 M_V^4} + \frac{F_V^2}{8 M_V^4} - \frac{f_\chi F_V}{2 \sqrt2 M_V^3},
\notag \\C_{72}^V &= \frac{G_V F_V}{8 M_V^4} -\frac{F_V^2}{8 M_V^4} + \frac{f_\chi F_V}{2\sqrt 2 M_V^3},\notag\\
C_{73}^V &= \frac{G_V F_V}{4 M_V^4} -\frac{F_V^2}{8 M_V^4} + \frac{f_\chi F_V}{\sqrt2 M_V^3},
\notag \\C_{74}^V &= -\frac{G_V^2}{4 M_V^4},\notag\\
C_{76}^V &= - \frac{G_V F_V}{8 M_V^4} +\frac{F_V^2}{16 M_V^4} - \frac{f_\chi F_V}{2\sqrt2 M_V^3},
\notag \\C_{78}^V &=  \frac{G_V F_V}{8 M_V^4} +\frac{F_V^2}{4 M_V^4} + \frac{f_\chi F_V}{2\sqrt2 M_V^3},\notag\\
C_{79}^V &=  - \frac{G_V F_V}{8 M_V^4} + \frac{F_V^2}{8 M_V^4} - \frac{f_\chi F_V}{2\sqrt2 M_V^3},
\notag \\C_{82}^V &=  - \frac{G_V F_V}{16 M_V^4} -\frac{F_V^2}{16 M_V^4} - \frac{f_\chi F_V}{4\sqrt2 M_V^3} - \frac{\fvdeux F_V}{\sqrt2 M_V^3},\notag\\
C_{83}^V &= \frac{3 G_V^2}{16 M_V^4} + \frac{f_\chi G_V}{2\sqrt2 M_V^3}-\frac{\sqrt2 \fvdeux G_V}{M_V^3},
\notag \\C_{87}^V &= \frac{F_V^2}{8 M_V^4},\notag\\
C_{88}^V &= - \frac{G_V F_V}{4 M_V^4} -\frac{f_\chi F_V}{\sqrt2 M_V^3},
\notag \\C_{89}^V &= \frac{F_V^2}{2 M_V^4} + \frac{G_V F_V}{4 M_V^4},\notag\\
C_{90}^V &= -\frac{f_\chi F_V}{\sqrt2 M_V^3},
\notag \\C_{92}^V &= \frac{F_V^2}{M_V^4},\notag\\
C_{93}^V &= -\frac{F_V^2}{4 M_V^4}\ .
\end{align}
In these formulas $n$ stands for the number of flavours and should be
set to $n=3$.

These results can be verified to agree with the ones obtained in
ref.\cite{kaiser} when
retaining the same resonance coupling constants as in our Lagrangian.
The correspondance in the notation between the coupling constants
appearing in our eq.\rf{efflag} and those in ref.\cite{kaiser} is as follows
\begin{align}
 \evm\,\frac{F_V^2}{2M_V^4}&=
\overline{\lambda}_6^{VV}+\overline{\lambda}^{SVV}
& \gvun\,\frac{F_V}{M_V^3}&= \overline{\lambda}_1^{SV}\notag \\
 \gvdeux\,\frac{F_V}{M_V^3}&= - \overline{\lambda}_2^{SV}
& \fchi\,\frac{F_V}{M_V^3}&=  \overline{\lambda}_1^{PV}\\
\fvdeux\,\frac{F_V}{M_V^3}&= - \overline{\lambda}_2^{PV}-\frac{1}{2}
\overline{\lambda}_1^{PV}
& \fvun\,\frac{F_V}{M_V^3}&=  \overline{\lambda}_3^{SV}\notag \ .
\end{align}
These relationships may be derived by making a field redefinition on the
scalar and pseudoscalar resonance fields used in ref.\cite{kaiser}
\be
S\to\tilde S + c_m\,\frac{\Chip}{M_S^2},\qquad
P\to\tilde P + i d_m\,\frac{\Chim}{M_P^2}\ .
\en
We note that the terms proportional to $f_\chi^2$  in
$C_{25}^V$ and $C_{26}^V$ which are generated in the $V-$formalism but not
directly in the $AT-$ formalism have not been considered in ref.\cite{kaiser}.

\subsubsection
{Resonance saturation versus experiment for $C_{61}$}\lblsec{c61}
Only one of the LEC's which appear in eqs.\rf{civ} (except for
$C_1$ and $C_4$) has actually been
determined from experiment. Let us consider the two-point correlator
of two vector currents
\bea\lbl{vvcorr}
&&i\int d^4x e^{ipx}\,\trace{ 0\vert T ( V_\mu^{ij}(x) V_\nu^{ji}(0)) \vert 0}
\\
&&\quad=  (p_\mu p_\nu -p^2 g_{\mu\nu})\,\Pi^{ij} (p^2)
+ g_{\mu\nu} p^2\, \Pi^{ij}_0(p^2)\nonumber
\ena
with
\be
V_\mu^{ij}(x)= \bar\psi^i(x)\gamma_\mu\psi^j(x)\ ,
\en
and then consider the difference
\be
\Delta\Pi = \Pi^{ud}(0)- \Pi^{us}(0)\ .
\en
The chiral computation of this quantity at order $p^6$ was first
performed in ref.\cite{durrkambor} and the result was confirmed and expressed
in terms of the canonical set of $O(p^6)$ LEC's in ref.\cite{ABT2pt}.
The chiral expansion involves no LEC at all at chiral order $p^4$ and
a single LEC at chiral order $p^6$, which is $C^r_{61}$.
Using finite-energy sum rule techniques, the value of
$\Delta\Pi$ can be determined from experiment\cite{durrkambor} (earlier
related calculations were performed in refs.\cite{golokamb1,maltman})
\be
\Delta\Pi_{exp}= 0.0203\pm 0.0032\ .
\en
This result translates into the following value for the $O(p^6)$ LEC
\be\lbl{c61exp}
C_{61}^r(m_\rho)= (1.24 \pm 0.44)\,10^{-3}\ {\rm GeV}^{-2}\ .
\en
On the other hand, our resonance saturation model, using the results
from eqs.\rf{civ} and the determination of the resonance parameters
discussed above, yields
\be
C_{61}^V = 2.10\,10^{-3}\ {\rm GeV}^{-2}\
\en
(using $F_V=F_\rho$, see eq.\rf{froexp})
which is in qualitative agreement with the experimental determination.

\subsection{Comparison between resonance saturation and the dispersive
representations}

We can now return to the $\pi K$ scattering amplitude and compute the
vector meson  contributions generated from the saturation of
LEC's $C_i$ as shown above \rf{civ}. We quote the result for the three
sub-threshold coefficients under consideration in this section,
\bea\lbl{cijciv}
&&\left.C^+_{20}\right|_{C_i^V}= \Big[
-{7\over8}\gvd {\mkd+\mpid\over\mvq}
+{3\over2}\gvd\evm {\mkd\over \mvq}\nonumber \\
&&\quad -{3\over\sqrt2} G_V\,{2\ghvun\mkd +\ghvdeux\mpid\over M_V^3}
\Big]{m_\pi^4\over F_\pi^4}\nonumber \\
&&\left.C^+_{01}\right|_{C_i^V}= \Big[
 2         \gvd {\mkd+\mpid\over\mvq}
-8         \gvd\evm {\mkd\over \mvq}\nonumber \\
&&\quad +8\sqrt2        G_V\,{2\ghvun\mkd +\ghvdeux\mpid\over M_V^3}
\Big]{\mpid\mkd\over F_\pi^4}\nonumber\\
&&\left.C^-_{10}\right|_{C_i^V}= \Big[
 3         \gvd {\mkd+\mpid\over\mvq}
-4         \gvd\evm {\mkd+2\mpid\over \mvq} \nonumber\\
&&\quad +4\sqrt2       G_V\,{2(\ghvun+\ghvdeux)\mkd
                    +(4\ghvun+\ghvdeux)\mpid\over M_V^3}
\Big] \nonumber\\
&&\ \quad \times {m_K m_\pi^3\over F_\pi^4}\ .
\ena
A comparison of the numerical results for the resonance saturated
part of these sub-threshold parameters between the vector field model
and the antisymmetric tensor model is performed in table 3. One can
see that the differences are substantial. In two cases even the sign of
the result is different.
\begin{table}
\centering
\begin{tabular}{|c||c|c|}\hline
               & V model\cite{bijdontpik} & AT model \\ \hline
$\left.C^+_{20}\right\vert_{C^i_V}$& $-$0.005&$-$0.010  \\
$\left.C^+_{01}\right\vert_{C^i_V}$& $-$0.27 & 0.30   \\
$\left.C^-_{10}\right\vert_{C^i_V}$& $-$0.11 & 0.21  \\ \hline
\end{tabular}
\caption{\sl Results on the $O(p^6)$ part
involving the $C_i^r$ LEC's of some sub-threshold coefficients,
using two different vector resonance saturation models of
these.}
\end{table}

One can perform a check of the resonance saturation model
in the following way.
Consider the set of sub-threshold coefficients which
can be written as sum rules with no subtractions. At this level,
it is easy to identify a particular resonance $R$ contribution : it suffices
to restrict the integration region to the neighbourhood of the resonance
mass and to restrict the sum over partial-waves to the one which corresponds
to the spin of the resonance. This is illustrated in fig. 2 which shows
the integrands (in both the $s$ and the $t$ channel) associated with the
coefficient $C^+_{20}$. In this case, the contribution from the vector
resonance can be isolated in the $s$ channel and the contributions from
the scalar resonances can be identified in both the $s$ and the $t$ channel.
Fig. 3 illustrates the situation for the coefficient $C^-_{10}$: in this case
the vector contribution appears in both the $s$ and the $t$ channels.

\begin{figure*}
\centering
\includegraphics[width=7cm]{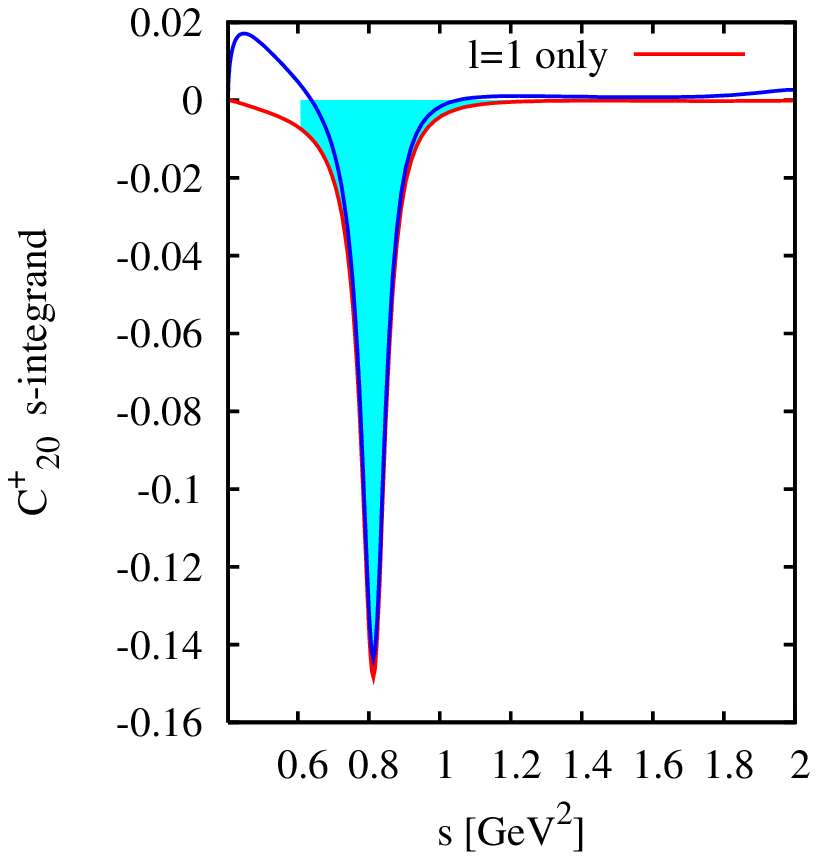}
\includegraphics[width=7cm]{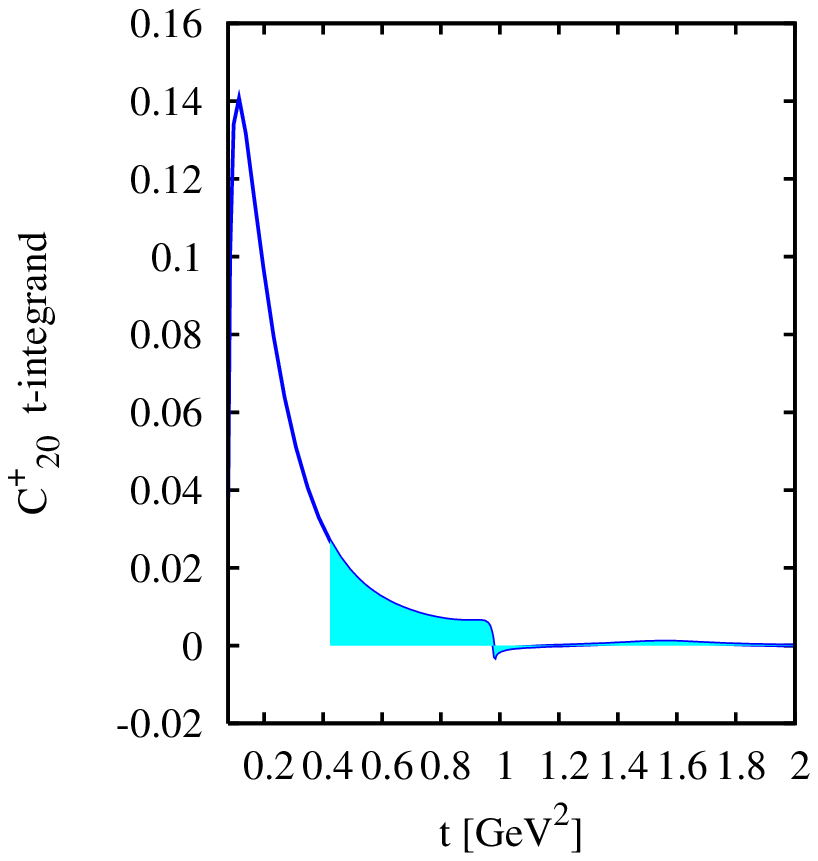}
\caption{\sl $C^+_{20}$
integrands: s-channel (left figure) and t-channel
(right figure). The shaded area on the left figure isolates the $K^*(890)$
resonance contribution and on the right figure the scalar $f_0(980)$ one. }
\end{figure*}

\begin{figure*}
\centering
\includegraphics[width=7cm]{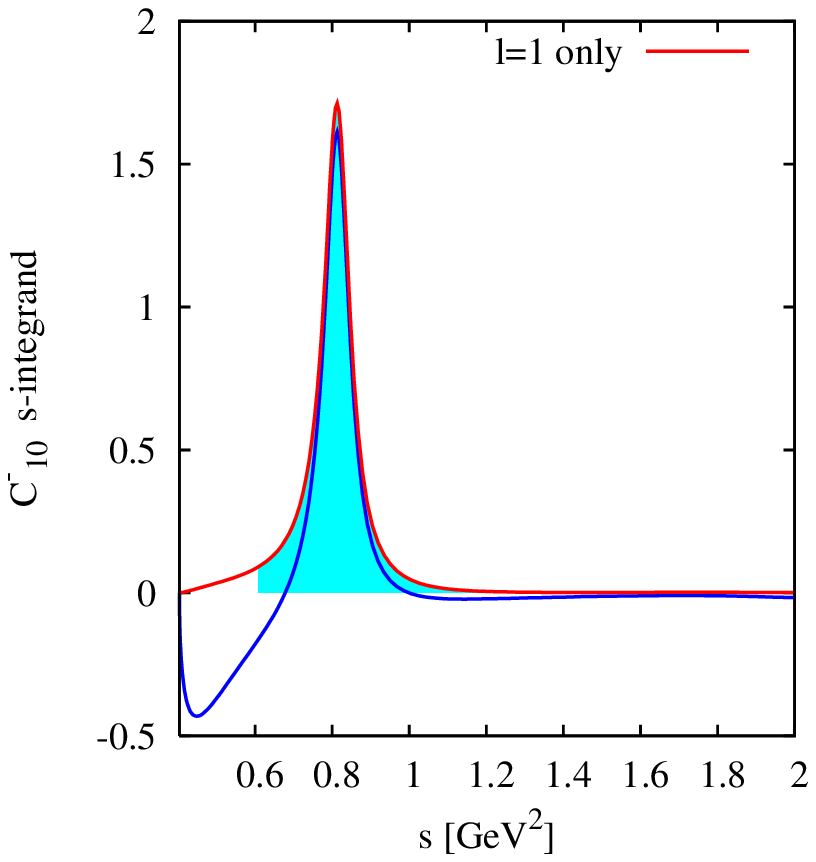}
\includegraphics[width=7cm]{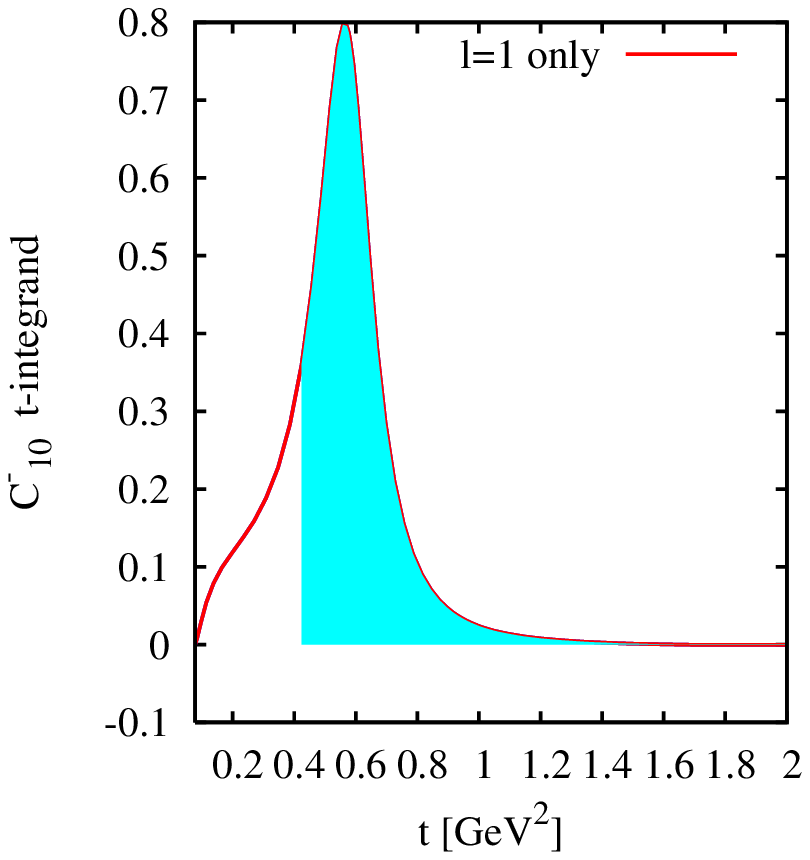}
\caption{\sl $C^-_{10}$
integrands: s-channel (left figure) and t-channel
(right figure). The shaded area on the left figure isolates the $K^*(890)$
resonance contribution and on the right figure the  $\rho(770)$ one. }
\end{figure*}

From the point of view of ChPT now,
we can split the contributions to a given sub-threshold coefficient $C_{ij}$
into one part,  $C_{ij}^{loop}$, which arises from loop diagrams
and one part $C_{ij}^{tree}$ which arises from  tree level diagrams.
The latter piece,
up to chiral order $p^6$,
involves terms linear in the $O(p^4)$ LEC's $L_i^r$, terms
which are quadratic in the $O(p^4)$ LEC's
and, finally, those which are linear in the LEC's $C_i^r$.
Both $C_{ij}^{loop}$ and $C_{ij}^{tree}$ depend on the chiral renormalization
scale $\mu$.
Let us assume that a proper scale $\mu$ exists such that $C_{ij}^{loop}$
corresponds to the low energy integration part of the coefficient $C_{ij}$
and $C_{ij}^{tree}$ to the higher energy part.
We can then make a check of our resonance saturation model by computing
$C_{ij}^{tree}$ using the reso\-nance-saturated values of the LEC's $L_i$
and $C_i$ and comparing the result with the dispersive integral calculation
in which the integral is computed over an energy range $E > E_0$. The
lower boundary of the
integration range should be somewhat below the resonance mass.
We will only consider the role
of the vector mesons here. In the resonance saturation model, we
correspondingly keep
the terms proportional to the coupling $G_V$.
The terms arising from the LEC's $C_i^r$ were shown in eq.\rf{cijciv}.
Upon using the resonance model of ref.\cite{egpr} and retaining the
contributions proportional to $G_V$ the terms which are linear or
quadratic in the LEC's $L_i^r$ yield
\be 
\begin{array}{l}
 \left.C^+_{20}\right\vert_{L+LL}=
-\dfrac{3}{8}\,\dfrac{G^2_V}{M_V^2}\left[
1- 8\dfrac{c_d c_m (\mkd-\mpid)}{F_\pi^2 M_S^2}\right]
\dfrac{\mpiq}{F_\pi^4}
\\
  \left.C^+_{01}\right\vert_{L+LL}=
2\, \dfrac{G^2_V}{M_V^2}\left[
1- 8\dfrac{c_d c_m (\mkd-\mpid)}{F_\pi^2 M_S^2}\right]
\dfrac{\mkd\mpid}{F_\pi^4}
\\
 \left.C^-_{10}\right\vert_{L+LL}=
3 \dfrac{G^2_V}{M_V^2}\left[
1- 8\dfrac{c_d c_m (\mkd-\mpid)}{F_\pi^2 M_S^2}\right]
\dfrac{\mk\mpit}{F_\pi^4}\ .
\end{array}
\en
The comparison, as discussed above, of the resonance saturation result
with the dispersive resonance calculation is performed in table 4.
The table shows that the results from the antisymmetric tensor model
for the relevant $C_i$'s when added to the contributions linear
and quadratic in the $L_i$'s compares rather well with the resonance
contributions as computed from the sum rules.

\begin{table}
\centering
\begin{tabular}{|c|c|c|c||c|}\hline
\          & L+LL    & (L+LL+C)$_V$  & (L+LL+C)$_{AT}$ & sum rule \\ \hline
$C_{20}^+$ &$-0.0065$& $-$0.012      & $-$0.017        & $-$0.017 \\
$C_{01}^+$ &0.439    &  0.17         &    0.74         &    0.66  \\
$C_{10}^-$ &0.185    &  0.08         &    0.40         &    0.40   \\ \hline
\end{tabular}
\caption{\sl Comparison between vector resonance contributions to three
subthreshold coefficients as computed from sum rules (last column) and
as computed from resonance saturation models of the LEC's. The second column
shows the contributions which are linear and quadratic in the LEC's $L_i$
while the third and fourth column show the additional effect
of the LEC's $C_i$ using the vector or the antisymmetric tensor model
respectively}
\end{table}
\subsection{A LEC combination with dominant vector contributions}
In general, the low-energy couplings get important contributions from the light
vector mesons and also from the light scalar resonances\cite{egpr}.
Accounting for the scalar contributions is made difficult by
several features. Firstly, the OZI rule is rather strong\-ly violated
in the scalar meson sector.
This induces a large number of parameters in the resonance
Lagrangian which cannot be determined unless some assumptions are made:
see e.g. ref.\cite{cirigscal} for a recent discussion and some
examples of such assumptions. A second difficulty is caused by the
presence of the wide scalars (the $\sigma$ or $\kappa$ mesons).
Interferences between the contributions
from the wide scalars and the narrow ones
lead to structures in the partial wave
amplitudes (see e.g. fig. 2 right)
which are not well approximated by computing tree level
diagrams from a resonance Lagrangian. For these reasons, it is useful to
try to identify specific combinations of LEC's which receive small
contributions from the scalar mesons.
We can generate one such combination by starting
from $\pi K$ sub-threshold coefficients and forming the following combination
\be\lbl{cns}
C_{NS} = C^+_{01} +{2\mk\over\mpi}\, C^-_{10}\ .
\en
Indeed, this quantity satisfies an unsubtracted dispersion relation and,
by construction, it receives no resonant S-wave contributions from either
the $t-$ or the $s-$channels.
The only
resonant contributions are from the $l\ge 1$ partial waves.
The s-channel integrand is shown in fig. 4
while the t-channel integrand is the same, up to a scale factor, as that
shown in fig. 3. Computing the integrals
we find the experimental value of this quantity
\be\lbl{cnsexp}
C_{NS} = 4.27 \pm 0.17\ .
\en
Using the chiral expansion for $C_{NS}$ one finds that
the following combination of $O(p^4)$ and $O(p^6)$ LEC's is involved,
\bea\lbl{leff2}
&& L^{eff}_2(\mu)= L^r_2
+(\mkd+\mpid)\left(-2C^r_4 + C^r_{10}\right.
-2 C^r_{12}+ 2 C^r_{22}
\nonumber\\
&&\quad \left.+2 C^r_{23}- C^r_{25}\right)
+(4\mkd+2\mpid)\left(C^r_{11}-2 C^r_{13}\right)\ .
\ena

According to the remarks made above, this combination of LEC's receives no
contributions from the scalar mesons corresponding to virtual exchanges
in the $\pi K$ scattering amplitude. It does, however, pick up contributions
from the scalars via tadpole-type diagrams\footnote{We thank Roland Kaiser
for pointing this out to us.}. Such contributions have been accounted for
in our approach via flavour symmetry breaking effects
with the exception, however,  of the LEC $C^r_{12}$
(and for the $1/N_c$ suppressed LEC's).
This LEC receives no contribution from the resonance Lagrangian terms
which we have considered.
Fortunately, direct determinations exist for $C^r_{12}$ based on the
scalar form factors with either $\Delta S=0$
(ref.\cite{bijdscalff}) or $\Delta S=1$ (ref.\cite{jopOp6}). The latter
determination seems more precise and gives a value in the range
$-0.6\le 10^4\,C_{12}^r(m_\rho)\le 0.6\ {\rm GeV}^{-2}\,$ which implies that
the corresponding contribution in eq.\rf{leff2} is negligibly small.

\begin{figure*}
\centering
\includegraphics{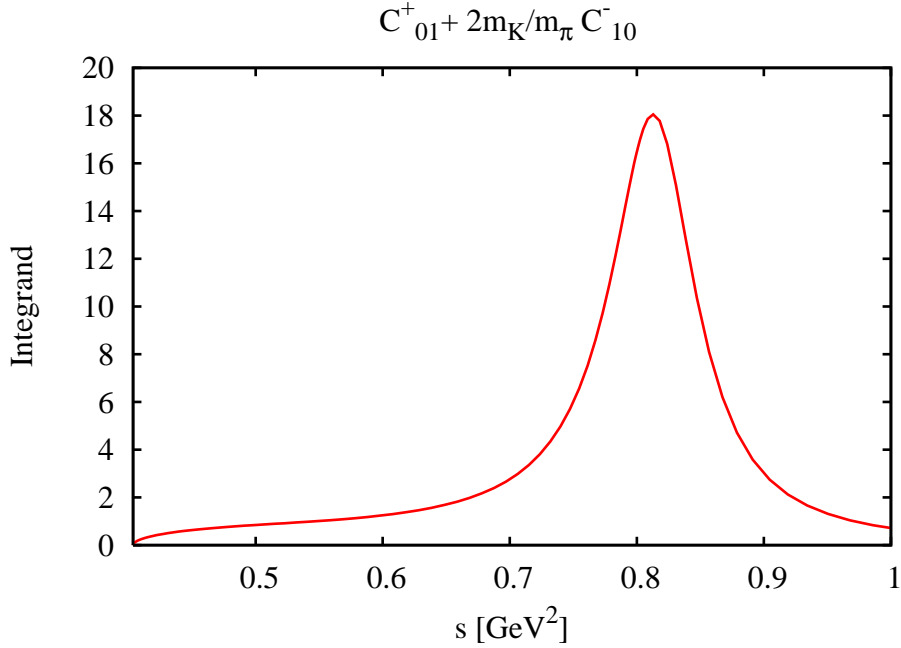}
\caption{\sl   Integrand in the s variable for the subthreshold coefficient
$C_{NS}$ defined in eq.\rf{cns}         }
\lblfig{fccombi}
\end{figure*}
We can determine the experimental value of $L^{eff}$ from the experimental
value of the combination $C_{NS}$ \rf{cnsexp} and its chiral expansion.
If we use the expansion up to order $p^4$ we find
\be
\left.L^{eff}_2(m_\rho)\right\vert_{p^4} \simeq 1.32\, 10^{-3}\ .
\en
This value agrees rather well with that found from the resonance saturation
model $L_2^V =1.2\,10^{-3}$ \cite{egpr}. If we now include the $O(p^6)$
correction in the chiral expansion we find
\be\lbl{leffexp}
\left.L^{eff}_2(m_\rho)\right\vert_{p^4+p^6}= (0.16 \pm 0.08)\,10^{-3}\ .
\en
We would like now to compare with the result from the resonance
saturation model also including $O(p^6)$ corrections which
has the following expression,
\bea
&& \left.L^{eff}_2\right\vert_V = {G_V^2\over 4\mvd}\Big\{
1+{\mkd+\mpid\over \mvd}\Big[
1-2\evm
\nonumber\\
&&\quad +2\sqrt2 {M_V\over G_V}(2\ghvun+\ghvdeux) \Big] \Big\}\ .
\ena
Numerically, using the results from sec.~3.4, one obtains
\be\lbl{leffV}
\left.L^{eff}_2\right\vert_V \simeq 2.04\,10^{-3}\ .
\en
Keeping in mind that the determination of the resonance Lagrangian
couplings is approximate (due, in particular, to the use of large $N_c$
type approximations), it is nevertheless clear that
the value of $L^{eff}_2$ obtained above \rf{leffV} from our resonance
saturation model differs quite substantially (by about a factor of ten)
from the experimental determination of $L^{eff}_2(\mu)$ when $\mu=m_\rho$.
\subsection{Discussion}
This problem cannot be attributed to the resonance model itself
since we have checked that the  results do
correspond, at least approximately, to the contribution
from the resonance region in the sum rule expression of $C_{NS}$
(the integrand is shown in fig. \fig{fccombi}).
It must therefore be concluded that the values of the LEC's
can fail to be dominated by the resonance contributions at
$O(p^6)$ with $\mu=m_\rho$.

One obvious possible reason for the failure of resonance saturation
is that the variation of the
LEC's as a function of $\mu$ at $O(p^6)$ can be much faster than it is
for the $O(p^4)$ LEC's. This is illustrated in
fig. \fig{muvar} which shows the behaviour of $L_2^{eff}$ as a function
of the scale.
The figure shows that, in fact, a scale $\mu_0$ does exist such that
resonance saturation of $L_2^{eff}$ is exact, but its value,
$\mu_0\simeq 0.45$ GeV is significantly smaller than $m_\rho$.

One must also keep in mind
that the renormalized coupling constants are obtained from the bare ones
by a minimal subtraction procedure. Their values thus depend both on the
regularization scheme and on the subtraction convention. The procedure
adopted in ChPT (based on dimensional regularization and modified minimal
subtraction) was shown to lead to natural values for the couplings at
order $p^4$. This, however, is not guaranteed to remain true at arbitrary
higher orders.
A remark is in order, finally, concerning the chiral expansion of the
quantity $L_2^{eff}$. In the resonance saturation model, the contribution
of order $p^6$ is rather large, amounting to a 50\% correction as compared
to the $O(p^4)$ one. At first sight, the situation seems to be
worse for $L_2^{eff}(\mu)$ : if we set $\mu=m_\rho$, the contribution
of order $p^6$ practically cancels that of order $p^4$. In this case,
however, the relative contributions strongly depend on the scale: if we
take $\mu\simeq 0.55$ GeV the $O(p^6)$ contribution will be much smaller
than the  $O(p^4)$ one, while if we take $\mu\simeq 0.45$ GeV the
relative contributions become similar to  those in the
resonance saturation model.

\begin{figure*}
\centering
\includegraphics{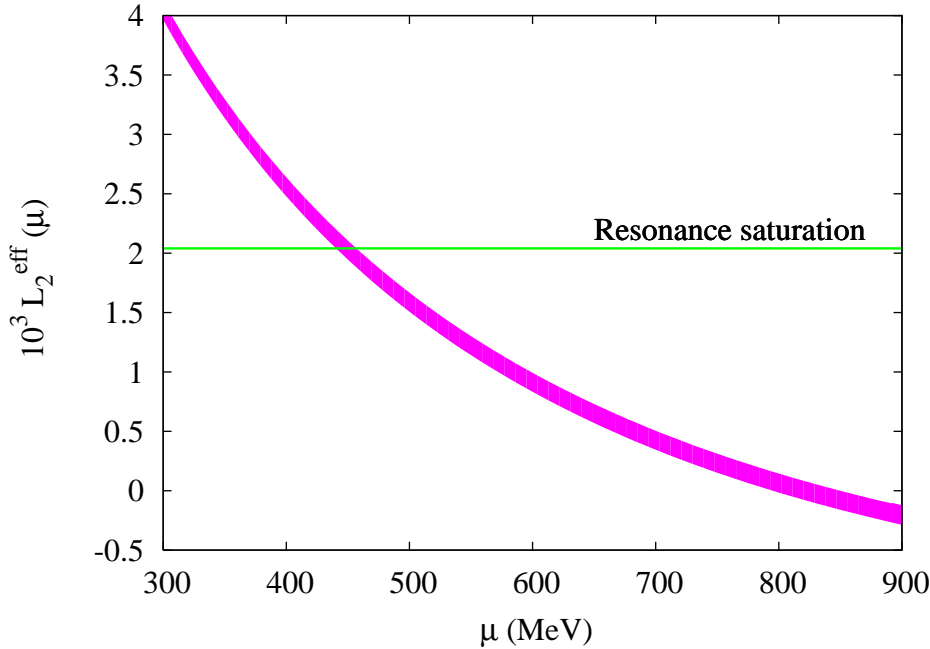}
\caption{\sl  Chiral coupling combination  $L^{eff}_2(\mu)$ (see \rf{leff2})
as a function of the scale $\mu$  compared with the vector meson
saturation result.        }
\lblfig{muvar}
\end{figure*}
\section{Summary}
Our goal was to extract some model independent informations about the
$O(p^6)$ chiral coupling constants, about which little is known at present,
and probe the validity, in this sector, of the idea of resonance
dominance.
We used input from the $\pi K$ scattering amplitude in the subthreshold
region derived from experimental data using dispersion relations. In this
way, we generated three constraints on the four LEC's $C_1$ to $C_4$
and three constraints on eleven LEC's among $C_5$ to $C_{22}$. These
are associated with chiral operators which involve one insertion of
the quark mass matrix.
In line with the earlier work of ref.\cite{bijdontpik} it appears natural,
assuming resonance dominance, to associate the values of these LEC's
with flavour symmetry breaking in the light resonance sector. In order
to implement this, we have considered a (vector) resonance Lagrangian
which is more general than the one used in ref.\cite{bijdontpik}. We determined
all the coupling constants in this Lagrangian from experiment, in a large $N_c$
spirit.
In principle, a more consistent approach to the determination of such couplings
is to appeal to asymptotic constraints\cite{eglpr}. In practice, the
two approaches usually give similar results and, furthermore, it is
often not possible to satisfy all the relevant asymptotic constraints
using a minimal number of resonances (e.g. \cite{knnyff,bijgamiz}).
Here, in order to test some of our estimates for the resonance content
of the LEC's, we have used unsubtracted sum rules in which one
restricts the integration range to the resonance region.

One of our initial motivations was to try to
understand the reason for a number of
significant discrepancies between the chiral $O(p^6)$ predictions
of ref.\cite{bijdontpik} for certain subthreshold expansion parameters
of the $\pi K$ amplitude and the dispersive results.
We found that improving the vector resonance Lagrangian does not
help in resolving these discrepancies. We made no attempt to improve
the scalar resonance Lagrangian but we identified a specific combination
of chiral LEC's which should be insensitive to that sector (beyond
the effect of generating flavour symmetry breaking). A clear outcome of
our analysis is that, if one sets the value of
the chiral scale $\mu$ equal to the $\rho$-meson mass, then
the value of this  combination of LEC's is not dominated by the resonance
contribution.
We have also encountered  examples for which resonance dominance
was reasonable, see sec.\sect{c61}.
This suggests that in parallel to the efforts which are pursued
in order to develop consistent resonance models (e.g. \cite{kaiser})
one should also try to obtain further
direct determinations of the LEC's $C_i$.

This result may be compared with the observation made in
the baryon sector of ChPT\cite{dono99} already at one loop.
In dimensional regularization, the
one-loop corrections to the baryon masses were found to be rather large
requiring, in order to compensate for that,
that the low-energy couplings be set to values which are unnaturally large.
The origin of the problem was traced to the regularization procedure and the
physical interpretation of the chiral scale $\mu$. One expects $\mu$ to
correspond approximately to a momentum cutoff in the loop integrals. The
authors of ref.\cite{dono99} show that this expectation can break down
when unequal mass particles propagate inside the loops.
As a possible cure to this problem they proposed
to use a regularization method
different from dimensional regularization.

\section*{Acknowledgments}

IPN is an unit\'e mixte de recherche du CNRS et de l'Universit\'e Paris-Sud 11
(UMR 8608).
Work supported in part by the EU RTN contract
HPRN-CT-2002-00311 (EURIDICE),
the European Community-Research Infrastructure Activity
under the FP6\quad ``Structuring the European Research Area"
program\-me (Had\-ron Phy\-sics, contract number RII3-CT-2004-506078)
and by Center for Particle Physics (project no. LC 527
of the Ministry of Education of the Czech Republic).

KK acknowledges warm hospitality at the Institut de Phy\-sique Nucl\'eaire
where  this work was initiated.

\end{document}